\documentclass[twocolumn,epsfig,pre]{revtex4}
\usepackage{graphics}
\usepackage{graphicx}
\usepackage{epstopdf}
\usepackage{amssymb}
\usepackage{amsmath}
\usepackage{hyperref}
\usepackage{latexsym}
\usepackage{color}
\usepackage{soul}
\begin{document}

	\renewcommand{\thefootnote}{\fnsymbol{footnote}}
	\renewcommand{\theequation}{\arabic{section}.\arabic{equation}}
	
	\title{Transport properties of polydisperse hard sphere fluid: Effect of distribution shape and mass scaling} 

	\author{Thokchom Premkumar Meitei}
	\author{Lenin S. Shagolsem}
	\email{slenin2001@gmail.com}
	\affiliation{Department of Physics, National Institute of Technology Manipur, Imphal, India} 
	
	\date{\today}

\begin{abstract}

\noindent A model polydisperse fluid represents many real fluids such as colloidal suspensions and polymer solutions. In this study, considering a concentrated size-polydisperse hard sphere fluid with size derived from two different distribution functions, namely, uniform and Gaussian and explore the effect of polydispersity and mass scaling on the transport properties in general. 
A simple analytical solution based on the Boltzmann transport equation is also presented (together with the solution using Chapman-Enskog (CE) method) using which various transport coefficients are obtained. The central idea of our approach is the realization that, in polydisperse system, the collision scattering cross section is proportional to a random variable \textit{z} which is equal to the sum of two random variables $\sigma_i$ and $\sigma_j$ (representing particle diameters), and the distribution of \textit{z} can be written as the convolution of the two distributions $P(\sigma_i)$ and $P(\sigma_j)$. 
The obtained transport coefficients are expressed as explicit function of polydispersity index, $\delta$, and their dependence on the nature of particle size distribution is explored. 
It is observed that in the low polydispersity limit, the transport coefficients are found to be insensitive to the type of size distribution functions considered.
The analytical results (for diffusion coefficients and thermal conductivity) obtained using Chapman-Enskog method and our simple analytical approach agrees well with the simulation. However, for shear viscosity, our analyical approach agress for $\delta \le 20\%$, while it agrees upto $\delta \approx 40\%$ with the result obtained using CE-method (in the limit $\delta \rightarrow 0$). Interestingly, the effect of scaling mass (i.e., mass proportional to the particle size and thus a random variable) produces no significant qualitative difference. 

\end{abstract}

\maketitle

\section{Introduction} \label{sec: intro}

Transport phenomena in fluids are consequences of gradients in flow speed, particle concentration or temperature throughout the bulk of the system. 
The flux $J$ and the gradient of the transported quantity $X$ is related through $J=C\nabla X$, where $C$ is the transport coefficient, representing Newton's, Fick's and Fourier's laws of transport process in fluids.\cite{Bird} The whole transport phenomena is governed by the motion of the constituent particles and the interaction between them. In all the thermodynamic states of low density fluids, the mean free path of collison is much greater than the diameter of constituent particles. The long mean free path makes the collison between these constituent particles a very rare event, and the binary collision is sufficient to describe the collision processes at this density regime thereby rendering the higher order collisions irrelevent. As the nature of interaction between the constituent particles determines the outcome of the collision process, transport coefficients provide a way to probe the forces between pair of constituent particles.\cite{her} The interatomic pair potential of argon atoms were determined for the first time from the available data of viscosity, and iterative inversion process was used to determine pair potential from the available data of viscosity for a monoatomic species in low density limit.\cite{Mait} This process was later extended to  mixture of polyatomic gases by Vesovic and Wakeham.\cite{VV} 

A polydisperse fluid, on the other hand, consists of an infinite number of species in which the parameters like shape, size, mass, energy, etc. form a continous distribution which can often be represented by some probability distribution function. While shape and size distribution is abundant in most of the real world systems,\cite{Duc} size distribution of a hard sphere (HS) fluid is considered in this study. Such a model system conveniently represents a real fluid as in the case of colloidal suspensions and polymer solutions. One of the basic characteristics of any soft matter system like colloids, polymer solutions and liquid crystals is polydispersity.\cite{Li,Wilding,Allar,Sollich,Maso} In such cases, the assumption that the system is uniform in these parameters or even that it consists of a finite number of species is often an oversimplification.\cite{Xu} Earlier studies on polydisperse systems revealed that various qualitatively new phenomena are observed as compared to monodispere counterpart. \cite{Dick,Pusey,Bol,Phan,Barrat,Rae,Bolh,Bart} Also, in various industrial processes which involve polydisperse granular materials, optimization of shape and size of such particles are necessary to improve their performance.\cite{Wea,Herr}

Dependence of transport properties on density for finite component as well as multi-component or polydisperse hard sphere systems were studied comprehensively over a wide range of system sizes through molecular dynamics simulations.\cite{Alder,Sig,Lue,Pie} Furthermore, many earlier work also suggested analytical formula for transport coefficients of such systems as a function of parameters like packing fraction, temperature etc.,\cite{Alder,Sig,Mei,Pie,Mur} but there are a few literature which provides a simple relation between transport coefficients and polydispersity index. 

The transport properties of fluids are quantitatively studied using the Boltzmann transport equation. Enskog and Chapman introduced a method to find the solution to this equation for a low density finite-species mixtures of gas.\cite{Fer} And for a dense hard sphere fluid, standard Enskog Theory (SET) and revised Enskog Theory (RET) are generally used to obtain the shear viscosity.\cite{Thorne,Tham,van} By taking polydisperse limit in the RET, Xu and Stell determined the shear viscosity coefficient for dilute as well as dense size polydisperse hard sphere fluids. However, in the study reported here, we consider an alternative and relatively simple approximation method to find the transport coefficients of a size-polydisperse hard sphere fluid as a function of polydisperity index using Boltzmann transport equation. To this end, we solve the Boltzmann equation for a single component system,\cite{Rief} and extend it to the polydisperse limit. 
Of all the possible distribution functions, we consider two very different functional form, namely, Gaussian and Uniform. The extent of size polydispersity is quantified through polydispersity index, $\delta=\omega/\mu$ with $\omega$ and $\mu$ as standard deviation and mean of the distribution, respectively. 
In this study, we aim to investigate the dependence of various transport coefficeints like diffusion coefficient, $D$, viscosity, $\eta$, and thermal conductivity, $\kappa$, on the polydispersity index, functional form of the distribution, and mass (i.e. polydisperse and scales as size or same mass). At the same time, we also calculate the aforesaid transport coefficients using Chapman-Enskog method for both uniform and Gaussian distributions. The analytical solutions are then compared with the results obtained from molecular dynamics simulations.\\
The remainder of the paper is organized as follows. In section~\ref{sec: model-description}, we present computer simulation details which is then followed by the description and solution of our model and Chapman-Enskog method in section~\ref{sec: Analytical solution}, and in section~\ref{sec: Results and discussion} the results are discussed. Finally, we conclude the paper in section~\ref{sec: conclusion}. 

\section{Model and simulation details} \label{sec: model-description}

\begin{figure}[h]
\includegraphics[width=0.4\textwidth]{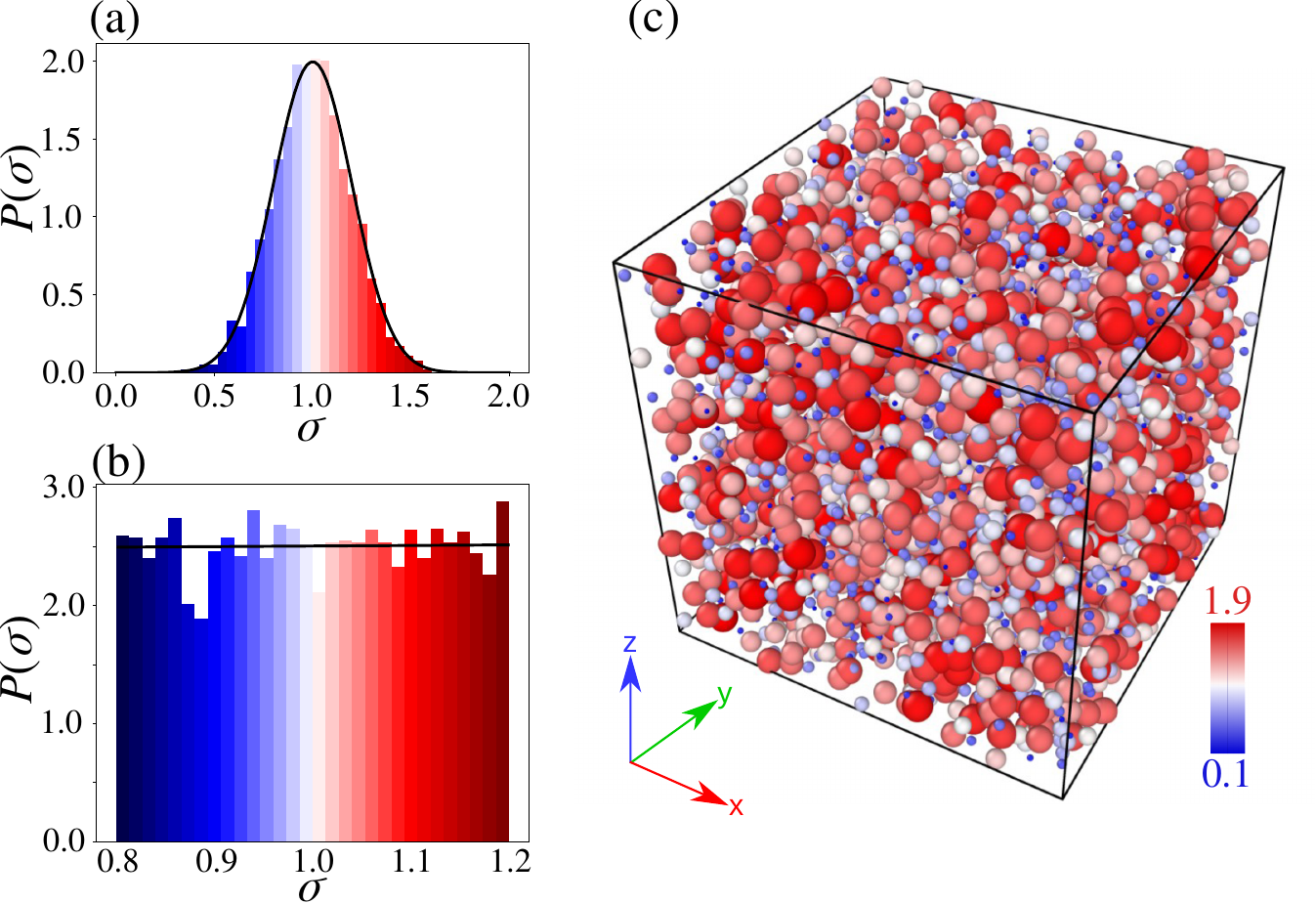}
\caption{Distribution function for particle size with (a) normal distribution of $\delta = 0.2$ (b) uniform distribution of $\delta = 0.11$. (c) Simulation snapshot of a  polydisperse HS fluid having uniform PSD with $\delta=0.11$.}
\label{fig: simu-snapshot}
\end{figure}

Coarse-grained molecular dynamics (MD) simulations in NPT ensemble are carried out for systems with $N = 5000$ size-polydisperse particles in three dimensions, see figure~\ref{fig: simu-snapshot}. The polydispersity in size is introduced by random sampling from two types of distributions, namely, uniform and Gaussian distribution functions.\cite{sheldon}
Uniform distribution, defined as 
\begin{equation}
P_u(\sigma)= \left\{ 
\begin{array}{l l}
\frac{1}{b-a} ~~~{\rm for}~ & a < \sigma < b ~\\
0 ~ & {\rm otherwise}
\end{array} \right.,
\label{eqn: uniform-dist}
\end{equation}
has mean and variance, 
\begin{equation}
\mu=(a+b)/2~~~{\rm and}~~~\omega^2=(a-b)^2/12~,
\label{eqn: uni-mean-vari}
\end{equation}
respectively. And Gaussian distribution is defined as
\begin{equation}
\label{eqn:normal-dist}
P_g(\sigma)=\frac{1}{\omega\sqrt{2\pi}}\exp\left[-\frac{1}{2}\left(\frac{\sigma-\mu}{\omega}\right)^2\right]~,
\end{equation}
with $\omega$ the standard deviation, and $\mu$ the mean. 
In this study, the polydispersity index vary in the range $\delta=0.1-0.5$ with $\mu=1$ for both types of ditributions. The masses of the particles are set to unity. At the same time we also consider systems whose mass is scaled according to the particle diameters for a meaningful comparision. The technical details of the simulation including the protocols to prepare the samples is detailed in appendix~\ref{app: simulation-details}.  
The diffusion constants are obtained directly by calculating the mean-square displacement from the particles trajectories, while the other transport coefficients such as viscosity and thermal conductivity are obtained using Green-Kubo (GK) method detailed in appendix~\ref{app: gk-method}. 
 
\section{Analytical solution} \label{sec: Analytical solution}

The statistical nature of a fluid in flow can be described by the Boltzmann transport eqaution,\cite{chap} an integro-differential equation, which gives the dynamics of the molecular distribution function and it can be exploited to determine the transport of heat, momentum or the particle concentration. The Boltzmann equation is written as 
\begin{equation} 
\frac{\partial f}{\partial t}+\vec{v}\cdot\frac{\partial f}{\partial \vec{r}}+\frac{\vec{F}}{m}\cdot\frac{\partial f}{\partial \vec{v}}=\int_{\vec{v_1}}\int_{\Omega'}(f'f_1'-ff_1)Vs\text{d}\Omega'd^3\vec{v_1}.
\label{eqn:BE}
\end{equation}
Here, $f(\vec{r},\vec{v},t)$ is the molecular distribution function of the system defined as 
$f(\vec{r},\vec{v},t)\text{d}^3\vec{v}\text{d}^3\vec{r}=$mean number of molecules whose centre of mass at time \textit{t} lies between $\vec{r}$ and$\vec{r}+\text{d}\vec{r}$ with velocity between $\vec{v}$ and $\vec{v}+\text{d}\vec{v}$. $\vec{F}(\vec{r},t)$ is the external force on the molecules. $\vec{v_1}$,$\vec{v_2}$ and $\vec{v_1}'$,$\vec{v_2}'$ are the initial and the final velocities of type 1 and type 2 particles before and after collision respectively such that $\vec{V}=\vec{v_1}-\vec{v_2}$ and $\vec{V}'=\vec{v_1}'-\vec{v_2}'$, $s(\vec{V}')\text{d}\Omega'=$ the number of molecules per unit time emerging after scattering with final realtive velocity $\vec{V}'$ with a direction in the solid angle range $\text{d}\Omega'$ about $\theta'$ and $\phi'$. And the following relations have been introduced for convenience
\begin{eqnarray}
f &=& f(\vec{r},\vec{v},t), ~f_1 = f(\vec{r},\vec{v_1},t) \\
f' &=& f(\vec{r},\vec{v}',t), ~f_1' = f(\vec{r},\vec{v_1}',t).
\end{eqnarray}

The standard method to find solutions to equation \ref{eqn:BE} is 
Chapman-Enskog method which uses perturbation series expansion approach,\cite{Fer,chap} where molecular distribution function is written as 
\begin{equation}
f=f^{(0)}+\xi f^{(1)}+\xi^2 f^{(2)}+...
\end{equation} 
with $\xi$ a very small parameter and $f^{(0)}$ is the equillibrium molecular distribution function and $f^{(1)},f^{(2)},...$ are correction terms. The viscosity, $\eta$, is related to the pressure tensor, $P_{zx}$, and velocity gradient through
\begin{equation}
\label{NL}
P_{zx}=\eta\left(\frac{\partial v_x}{\partial y}\right)~,
\end{equation}
where pressure tensor, in terms of \textit{f}, is  
\begin{equation}
P_{zx}=m\int\text{d}^3\vec{v}fU_zU_x~
\end{equation}
with $U_i=v_i-v'_i$. For instance, in zeroth order perturbation, \textit{f} follows Maxwell distribution and the viscosity of a monodisperse HS fluid is obtained as\cite{Fer,chap}
\begin{equation}
\label{eqn:eta}
\eta=\frac{5\sqrt{\pi}}{16}\frac{(mkT)^{1/2}}{\Omega_0}~,
\end{equation} 
with $\Omega_0$ the total scattering cross section (between two particles of sizes $\sigma_i$ and $\sigma_j$) give by 
\begin{equation}
\Omega_0=\pi(\sigma_i+\sigma_j)^2~.
\label{eqn:omega}
\end{equation}

\begin{figure}[!h]
	\includegraphics[width=0.8\linewidth]{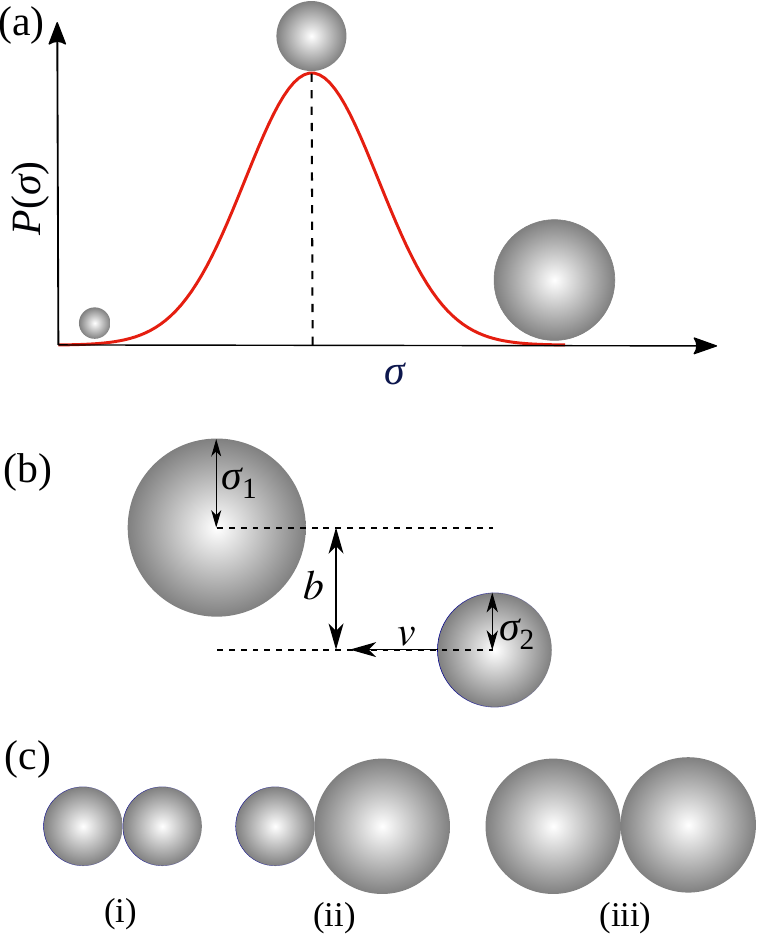}
	\caption{(a) Particle size distribution(PSD) of the HS fluid represented by normal ditribution. (b) Binary collision of two hard spheres of radii $\sigma_1$ and $\sigma_2$ (diameter of two randomly selected particles) with impact parameter \textit{b} (c) Representation of binary collisions of various possible pairs of hard spheres with radii represented by the random variables $\sigma$ having the probability distribution $P(\sigma)$.}
	\label{fig:random collision}
\end{figure}

In this work, we extend the zeroth order perturbation of Chapmann-Enskog (CE) solution to the polydispersity limit (using a relatively simple approach detailed below) and compare with the CE solution considering first order perturbation, also check the results obtained against the simulation data.\\


Consider a size polydisperse HS system with size distribution $P(\sigma)$ i.e.~the size of an $i^{\rm th}$ particle $\sigma_i$ is drawn randomly from the distribution and thus a random variable. The central idea of our approach is the realization that, mathematically, $\Omega_0=\pi z^2$ with $z=(\sigma_i+\sigma_j)$ represents another random variable, where the distribution of $z$ is given by the convolution of the two distributions $P(\sigma_i)$ and $P(\sigma_j)$, i.e., 
\begin{equation} 
Q(z)=\int P(z-\sigma_i)P(\sigma_i)\text{d}\sigma_i~,
\label{eqn: convolution-defn}
\end{equation} 
and thus the average total scattering cross section  
\begin{align} 
\Omega_0 = \pi \overline{z^2} = \pi\int z^2Q(z)\text{d}z~.
\label{eqn:cs}
\end{align} 
Thus, in this simplistic approach, calculation of various transport coefficients reduces to solving the above integral (eqn.~\ref{eqn:cs}) and use zero order perturbation result (eqn.~\ref{eqn:eta}). Below we consider the case of both uniform and normal distribution functions. 


\subsection{Uniform PSD}
\label{viscosity_uniform}
For uniform distribution (see eqn.~\ref{eqn: uniform-dist}), the convolution integral (eqn.~\ref{eqn: convolution-defn}) gives 
\begin{equation}
Q_u(z)= \left\{ 
\begin{array}{l l}
(z-2a)/(b-a)^2 ~~~{\rm for}~ & 2a<z<a+b  ~\\
(2b-z)/(b-a)^2 ~~~{\rm for}~ & a+b<z<2b  ~\\
0 ~ & {\rm otherwise,}
\end{array} \right.
\end{equation}
and thus   
\begin{eqnarray}
\bar{z^2}&=&\int_{0}^{\infty}z^2 Q_u(z)\text{d}z\notag\\
&=&\int_{2a}^{a+b}z^2 \frac{(z-2a)}{(b-a)^2}\text{d}z
+\int_{a+b}^{2b}z^2 \frac{2b-z}{(b-a)^2}\text{d}z\notag\\
&=&\frac{7a^2+7b^2+10ab}{6}
\label{eqn:uni1}
\end{eqnarray}
from which mean scaterring cross section can readily be obtained. Using eqns.~\ref{eqn: uni-mean-vari} we can express $\overline{z^2}$ in terms of polydispersity index, $\delta=\mu/\omega$, as  
\begin{align}
\overline{z^2}=2\mu^2(2+\delta^2)~.
\label{eqn: uni3}
\end{align}
Thus, an approximate expression for viscosity (using eqn.~\ref{eqn:eta}) for size-polydisperse HS system reads  
\begin{equation}
\frac{\eta}{\eta_0}=\frac{1}{1+\frac{1}{2}\delta^2} \approx 1-\frac{1}{2}\delta^2 + O(\delta^4)
\label{norm_eta:uni}
\end{equation}
with $\eta_0$ corresponds to $\eta$ for $\delta=0$. 
Moreover, the viscosity is related to diffusion coefficient $D$ through Stokes-Einstein (SE) relation\cite{Julia}, i.e., 
\begin{equation}
D=\frac{k_BT}{3\pi\eta d}
\label{eqn: SE}
\end{equation}
with $k_BT$ thermal energy, and $d$ diameter of the diffusing particle. Thus, an approximate expression of diffusion coefficient reads  
\begin{equation}
\frac{D}{D_0}\approx 1+\frac{1}{2}\delta^2 + O(\delta^4)
\label{eqn: D/D_0 uni}
\end{equation}
with $D_0$ represents the diffusion coefficient when $\delta=0$. In the following, we discuss the result for Gaussian distribution. 


\subsection{Gaussian PSD}
\label{viscosity_gaussian}
For Gaussian distribution (see eqn.~\ref{eqn:normal-dist}) we get 
\begin{eqnarray}
\overline{z^2}&=&\int_{0}^{\infty}z^2Q_n(z)\text{d}z\notag\\
			  &=&\int_{0}^{\infty}\frac{z^2}{\sqrt{4\pi\omega^2}}\exp\left[{-\frac{(z-2\mu)^2}{4\omega^2}}\right]\text{d}z~,
\label{eqn:norm1}
\end{eqnarray}
where $Q_n(z)$ represents convolution of the distribution $P_n(\sigma)$ with itself. Substituting $u$ for $(z-2\mu)$, the above equation reads 
\begin{equation}
\overline{z^2}=\frac{1}{\sqrt{4\pi\omega^2}}\int_{-2\mu}^{\infty}(u+2\mu)^2\frac{1}{\sqrt{4\pi\omega^2}}\exp\left({-\frac{u^2}{4\omega^2}}\right)\text{d}u
\label{eqn:norm3}
\end{equation}
and gives the following solution  
\begin{align}
\overline{z^2}=2\mu^2\left[(1+\frac{1}{2}\delta^2)\left(1+\text{erf}(1/\delta)\right)+\frac{1}{\sqrt{\pi}}\delta\exp(-1/\delta^2)\right].
\label{eqn: norm5}
\end{align}
In the limit $\delta\rightarrow 0$, $\exp(\cdot)\rightarrow 0$ and $\text{erf}(1/\delta)\rightarrow 1$, and hence $\overline{z^2}\approx 4\mu^2$ (which is same as the uniform distribution, see eqn.~\ref{eqn: uni3}, as expected). On the other hand, for $\delta >> 1$, the above equation~\ref{eqn: norm5} reads 
\begin{equation}
\overline{z^2}\approx 2\mu^2\left[(1+\frac{1}{2}\delta^2)+\frac{1}{\sqrt{\pi}}\delta\exp(-1/\delta^2)\right]~.
\label{eqn:norm6}
\end{equation}

Now, the expression of viscosity (eqn.~\ref{eqn:eta}) for the Gaussian distribution becomes
\begin{equation}
\label{eqn:normfinal} 
\frac{\eta}{\eta_0}=\frac{2}{(1+\frac{1}{2}\delta^2)(1+\text{erf}(1/\delta))},
\end{equation}
where $\eta_0$ corresponds to $\eta$ for $\delta=0$. And using the SE relation (eqn.~\ref{eqn: SE}) we obtain the diffusion coefficient as 
\begin{equation}
\label{eqn: D/D_0}
\frac{D}{D_0}\approx \frac{1}{2}\left(1+\frac{1}{2}\delta^2\right)\left(1+\text{erf}(\frac{1}{\delta})\right)
\end{equation}
where $D_0$ is the diffusion coefficient corresponding to $\delta=0$. Here, we assume the particle size $d\approx\mu$ (in the SE relation). Note that the expressions of $\eta$ and $D$ obtained here for the Gaussian PSD (see eqns.~\ref{eqn:normfinal} and \ref{eqn: D/D_0}) is same as that of the Uniform PSD (see eqns.~\ref{norm_eta:uni} and \ref{eqn: D/D_0 uni}) in the limit $\delta \rightarrow 0$ and thus it is expected that, for very small size polydispersity, the response of the system is insensitive to the shape of PSD. \\

\begin{figure}[!h]
	\includegraphics[width=0.9\linewidth]{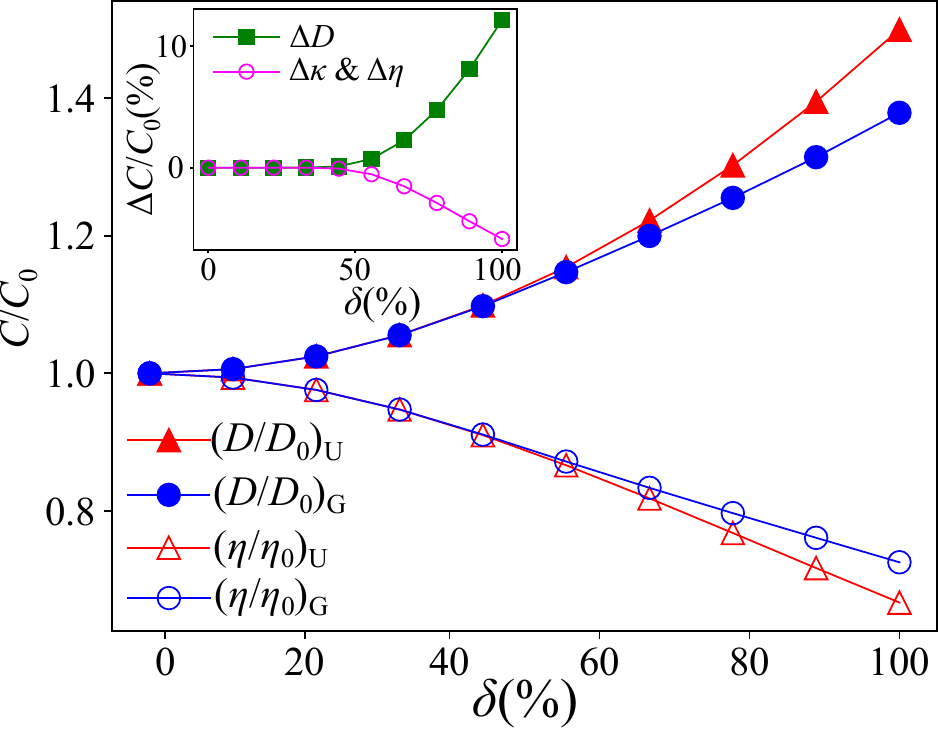}
	\caption{Theoretical normalised transport coefficients ($C/C_0$) [equation \ref{eqn: D/D_0 uni} and \ref{eqn: D/D_0} for $D/D_0$, \ref{norm_eta:uni} and \ref{eqn:normfinal} for $\eta/\eta_0$ and \ref{eqn: l/l_0} and \ref{norm: l/L_0} for $\kappa/\kappa_0$] as a function of the polydipsersity index ($\delta$). Here, all the fitting parameters are equated to unity. The  difference between the patterns of normal and uniform PSD is slightly larger for large polydispersity index. (Inset) The difference between normalized transport coefficients of uniform and normal PSD is negligible upto $\delta=0.5$ and the maximum value of the difference is less than 15\% at $\delta=100\%$. Hence, as real polydisperse fluids are in the small and mid-range polydispersity index, the nature of transport coefficient can be assumed to be same for fluids with uniform and normal PSDs.}
	\label{fig:theo}
\end{figure}

In figure~\ref{fig:theo}, we display the behavior of transport coefficients obtained for both Gaussian and uniform distributions. It is clear that the difference between the values of normalised transport coefficients of uniform and normal PSD is less than 0.001\% upto polydispersity index of $\delta=0.5$. This difference increases to 15\% at  $\delta=1$. However, to our knowledge, the polydispersity lies below $\delta=0.5$ (50\%) for real polydisperse systems\cite{elsana}, and hence, we can assume that the transport coefficients of such polydisperse fluids are independent of the type of PSDs under this study.  \\



\subsection{Viscosity using Enskog method}

The viscosity for a size-monodisperse HS fluid in the low density limit (using Enskog method~\cite{chap}) is given by 
\begin{equation}
\eta_e=\eta_0b_2\rho\left(\frac{1.016}{Z-1}+0.8+0.7737(Z-1)\right)
\label{eqn: Enskog}
\end{equation}
where $b_2=2\pi\sigma^3/3$ is the second varial co-efficient of the HS fluid, $Z$ is the compressiblity factor and $n_0$ is the viscosity of HS fluid in the infinite dilute limit.
Now, for a size-polydisperse system with distribution $P(\sigma)$, following the method outlined in reference~\cite{Xu}, the viscosity is given by 
\begin{equation}
\eta = \frac{5}{16\bar{\sigma}^2}\sqrt{\frac{mk_BT}{\pi}}p_\eta~.
\label{eqn: xu}
\end{equation}
Here, mass of each particle is assumed to be same and the parameter $p_\eta$ is obtained from the following relation, i.e.,
\begin{equation}
p_\eta = 48 p_1^2\frac{d_0(d_2-2)-d_1^2}{\Delta}
\label{eqn: p_eta}
\end{equation}
\begin{align}
{\rm with}~~
\Delta = &(d_2-2)(d_2-d)^2 + d_0d_3^2 + d_1^2d_4\notag \\ &-d_0d_4(d_2-0)-2d_1d_3(d_2-4)~,\\
\text{and}~~
d_j = &\int_0^\infty\frac{P(\sigma)\sigma^j}{\sigma^2+2p_1\sigma+p_2}\text{d}\sigma.
\label{eqn: Enskog 1}
\end{align}
Here, $p_0=1$, $p_1=\bar{\sigma}$ and $p_2=\bar{\sigma}^2+\text{var}(\sigma)$ are zero, first, and second moment, respectively, of the size distribution function.  
For the distribution functions considered in this study, we solve numerically the above set of equations (\ref{eqn: xu}-\ref{eqn: Enskog 1}) and obtain the value of $\eta$ at a given value of variance (or $\delta$). An approximate expression is also obtained in the limit of $\delta\rightarrow 0$, e.g., for uniform PSD (eqn.~\ref{eqn: uniform-dist}) with $\mu=1$, the expression for viscosity becomes 
\begin{equation}
\label{eqn: D_CE}
\frac{\eta}{\eta_0} =1 + \frac{1}{4}\delta^2 - \frac{1}{32}\delta^4 - \frac{169}{1280}\delta^{6} + O(\delta^{8})~.
\end{equation}
Different functional form of $P(\sigma)$ should lead to the similar functional form of $\eta$ at relatively small polydispersity index $\delta$. In the folowing section, we present comparison of the analytical results with that of the simulations data. 


\begin{figure}[h]
	\includegraphics[width=0.9\linewidth]{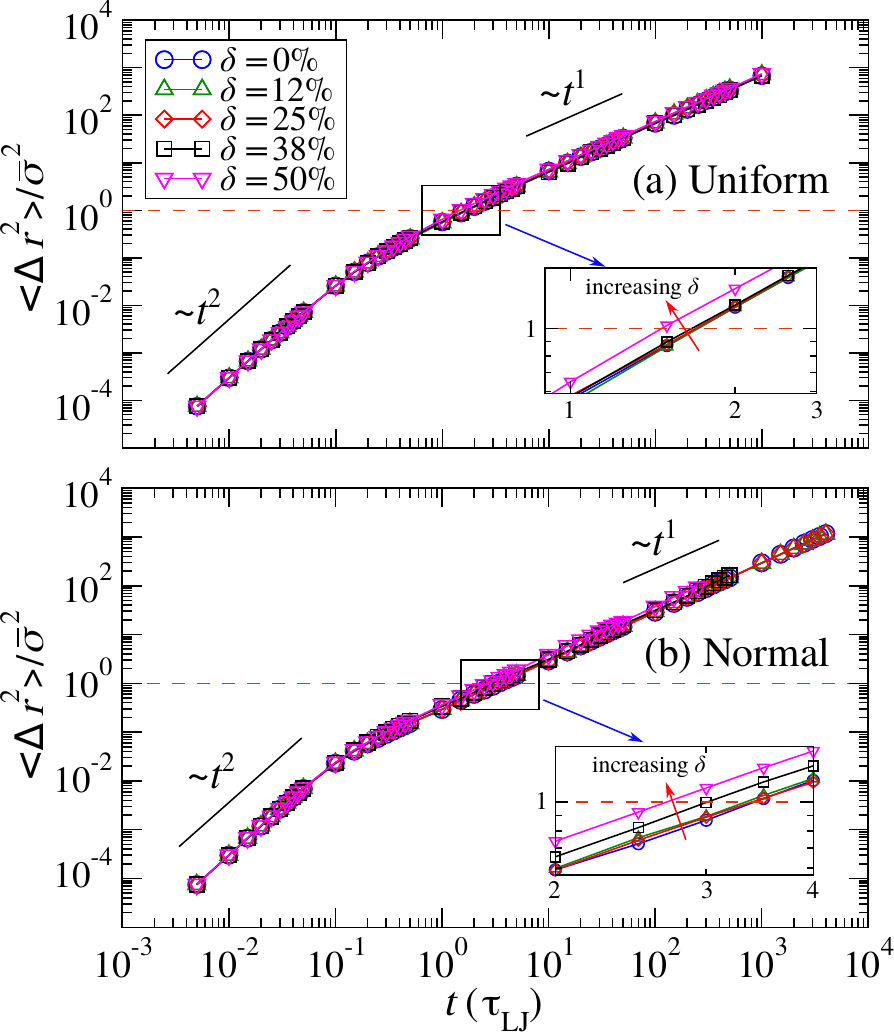}
	\caption{MSD curve of polydisperse HS fluids (where all masses are same) having Gaussian (above) and uniform (below) PSDs with mean 1 but of different polydispersity indices($\delta$). The early time behaviour indicates $t^2$ dependence while the late time regime shows normal Brownian diffusion ($\langle \Delta r^2\rangle \sim t^1$). Inset: The mean structural relaxation times ($\tau_\alpha$) of the systems under study shifts towards left with increasing polydipsersity index ($\delta$).}
	\label{fig: msd}
\end{figure}

\section{Comparison with simulation} \label{sec: Results and discussion}

\subsection{Diffusion Coefficients} \label{subec: D}

\begin{figure}[h]
	\includegraphics[width=1\linewidth]{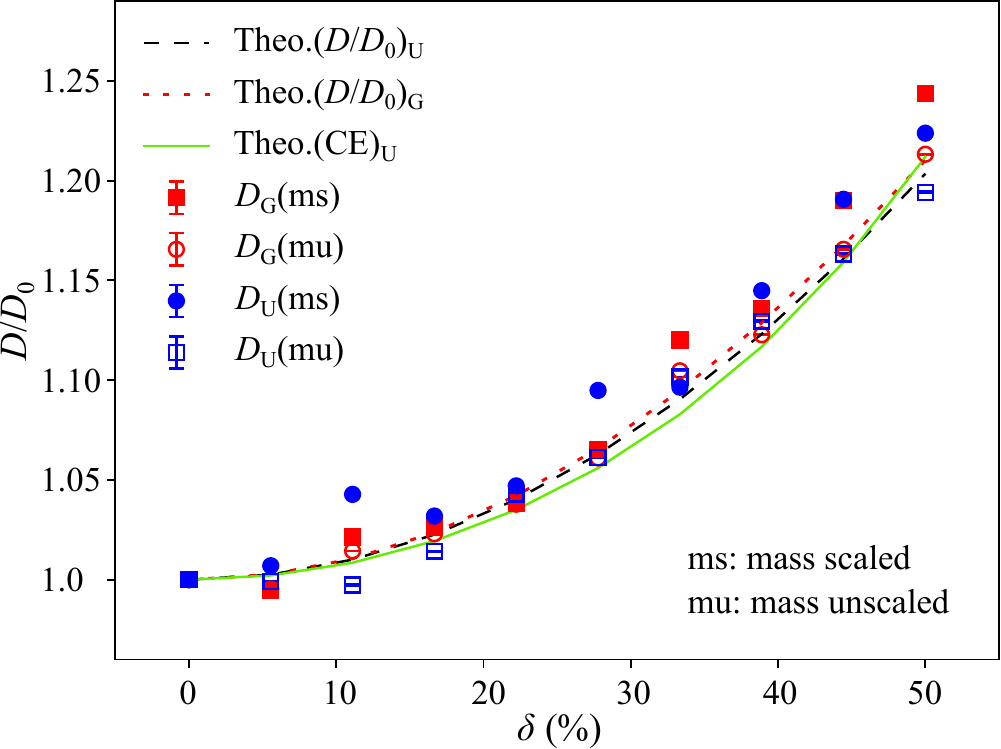}
	\caption{Normalised diffusion coefficient  of polydisperse HS fluids with different PSDs. The filled symbols corresponds to the system in which masses of the particles have been scaled proportional to the cube of their diameters ,while the unfilled symbols, to the system in which all the masses are set to unity. Dashed lines represents eqn \ref{eqn: D_U}, dotted line , eqn \ref{eqn: D_G} and the solid line is Chapman-Enskog method(eqn \ref{eqn: D_CE}). For the the curve-fitting process, we have introduced unitless fitting paratemeters as coefficients of the terms involving $\delta^2$ as discussed further in appendix~\ref{cf}. And data used for fitting are obatined by MD simulations of the mass unscaled systems.}
	\label{fig: Diffusion comp}
\end{figure}

A direct means to probe the particle dynamics is through the mean-square displacement (MSD) during time $t$ which we obtained from the particle trajectories. Mathematically, MSD is calculated by using the relation
\begin{equation}
\label{eqn: msd}
\langle \Delta r^2 (t) \rangle =\frac{1}{N}\sum_{i=1}^N \left[ r_i(t)-r_i(0)\right]^2~,
\end{equation}
where $r_i(t)$ is the position of $i^{\rm th}$ particle at time $t$, and $N$ the number of particles. The slope of the MSD curves gives the diffusion coefficient, $D$, i.e.,  
\begin{equation}
\label{eqn: msd_t}
D = \langle\Delta r^2\rangle / 6t^\alpha~,
\end{equation}
where the exponent $\alpha$ determines the qualitative nature of the motion. With $\alpha=1$ the particles exhibit normal diffusion, while for $\alpha>1$ (or $\alpha<1$) it is super-diffusive (or sub-diffusive). In figure~\ref{fig: msd}, we display MSD curves for both uniform and Gaussian size-polydisperse systems at different values of polydispersity index. In both types of systems, irrespective of the value of $\delta$, inertial (i.e. $\langle\Delta r^2\rangle \sim t^2$) as well as diffusive (i.e. $\langle\Delta r^2\rangle \sim t^1$) regimes are observed. 
As shown in figure~\ref{fig: Diffusion comp}, the diffusion coefficient, $D$, increases with increasing $\delta$, and the simulation data is fitted reasonably well with the equation derived. 
The increasing trend may be understood qualitatively form the behavior of mean relaxation time $\tau_\alpha$ (see figure \ref{fig: msd} inset). Systems with higher $\delta$ have lower value of relaxation time $\tau_\alpha$, and hence, higher value of \textit{D}. The effect of mass scaling is not significant though the systems with scaled mass have slightly higher value of normalised diffusion coefficient. Particles with diameter smaller than the mean value have dominant contribution on the diffusion coefficient of the polydisperse system. On scaling the masses of the particles proportional to the cube of their diameters, the masses of these smaller particles are lower than the case where mass is assumed to be unity for all species. Hence, the mass scaled system will be more diffusive as compared to the mass unscaled system.

\begin{figure}[h]
	\begin{center}
		\includegraphics[width=0.9\linewidth]{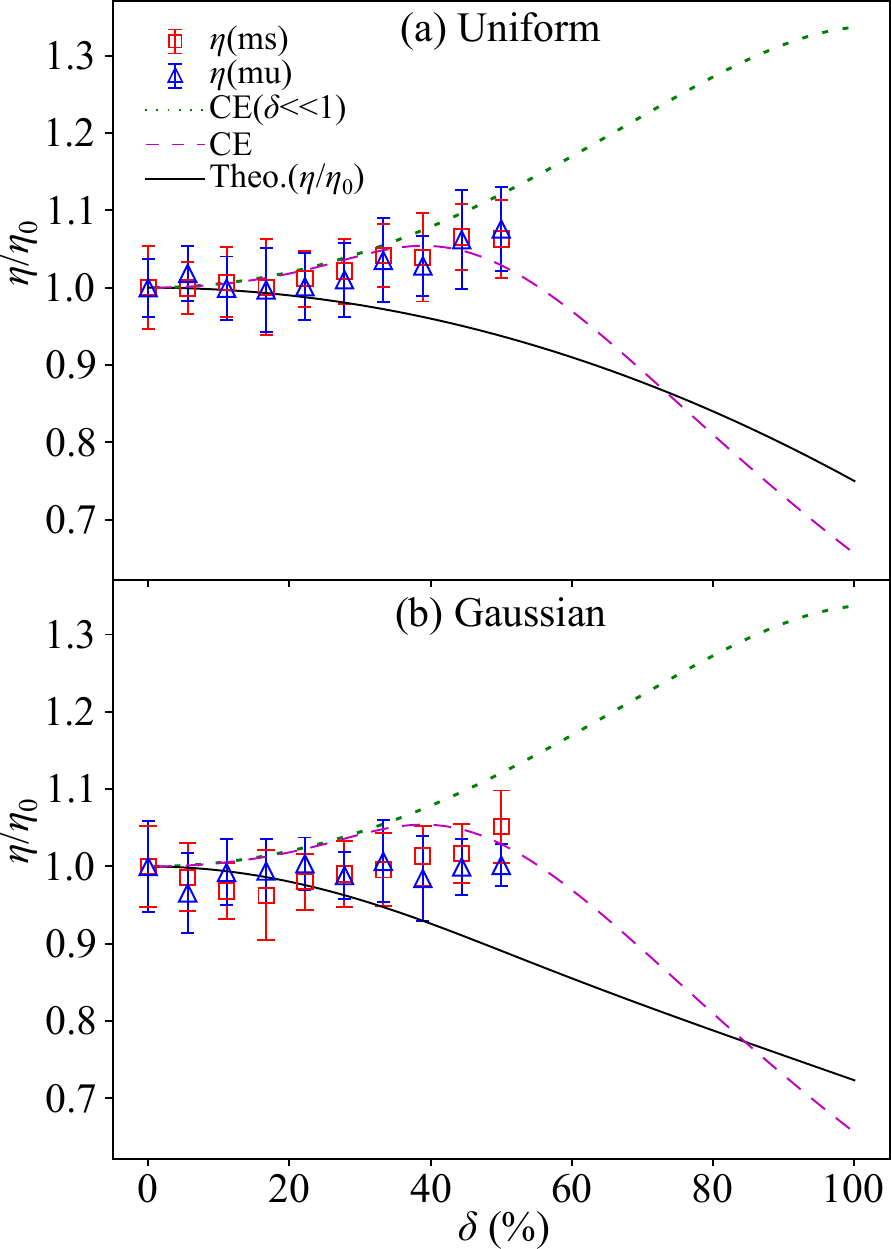} 
		\caption{Normalized viscosity for HS fluids with uniform(above) and Gaussian(below) PSDs. The solid lines represent our analytical equations~\ref{norm_eta:uni} and \ref{eqn:normfinal},the dotted lines is eqn \ref{eqn: D_CE}(by introducing a fitting parameter as a coefficient of $\delta^2$ and its value is 1.2) while the dashed lines are plotted according to the procedure described in the reference~\cite{Xu} [eqns~\ref{eqn: xu}-\ref{eqn: Enskog 1}]. Here, our analytical aproximation diverges form the MD simulations beyond $\delta=30\%$. The nature of dependence of viscosity on $\delta$ is similar for both PSDs. Here, the solid and the dashed lines are plotted  without using fitting parameters}
		\label{fig:norm viscosity}
	\end{center}
\end{figure}

\subsection{Viscosity}
\label{subec: V}

Shear viscosity of a fluid is a measure of its tendency to transport momentum in a direction normal to that of velocity, which manisfests as the resistance of the fluid to being sheared or the flow with varying  velocities within the fluid itself. Mathematically, it is defined by Newton's law of viscosity as the ratio of the shearing force per unit area to the velocity gradient in the fluid~[equation~\ref{NL}]. The Green-Kubo method is used for the calculation of the shear viscosity of HS systems with different values of polydispersity indices through MD simulations.

The dependence of normalised vicosity on the polydispersity index($\delta$) is shown in the figure~\ref{fig:norm viscosity} for systems with uniform as well as Gaussian PSDs. Our analytic solution predicts decrease of $\frac{\eta}{\eta_0}$ with increasing polydispersity index ($\delta$) in a quadratic fashion~[equations \ref{eqn:normfinal}, \ref{norm_eta:uni}]. This prediction is accurate in comparison to the simuations for low polydispersity upto $\delta=20\%$. However, beyond this point  the simulations show that $\frac{\eta}{\eta_0}$ increases slowly to attain a maximum value which is qualitatively  in agreement with Chapmann-Enskog method for polydisperse HS systems.~\cite{Xu} The systems with scaled mass and those with unscaled mass follow similar trend, but the system with scaled mass is slightly viscous for most of the $\delta $ values though the difference is negligible~(figure \ref{fig:norm viscosity}(a)). Again, as in the case of diffusion constant (section~\ref{subec: D}), the normalised transport coefficient $\eta/\eta_0$ is insensitive to the type of the particle size distribution(PSD) functions though it shows strong dependence on $\delta$.  Hence, the simplified relations for shear viscosity provided here will be useful for size polydisperse fluids of all types of particle size distribution(PSD) functions in low polydispersity limit ($\delta \leq 20\%$). In the following we discuss the behavior of thermal conductivity.


\subsection{Thermal conductivity}
\label{subsec: kappa}
In fluids, heat conduction can be considered as the process of the transport of molecular energy. The rate of conduction is dependent on the thermal conducitivity($\kappa$) of the fluid, which is defined by Fourier's law of heat conduction as the ratio of the temperature gradient to the heat flow per unit area. For an isotropic system, the relation between heat flux vector($\vec{q}$) and temperature gradient  is given by 
\begin{equation}
\label{FL}
\vec{q}=\kappa\nabla T~.
\end{equation}
For a monoatomic gas with low density, Chapmann-Enskog\cite{JO} provided a rigorous way to calculate thermal conductivity at temperature \textit{T}, rewritten for a HS system as 
\begin{equation}
\label{CE kappa monodisperse}
\kappa=\frac{125R}{64}\frac{\sqrt{\pi m k_BT}}{\pi\sigma^2}
\end{equation}
with \textit{R}= gas constant. 
This result can be extended to the limit of poldisperse HS following the same procedure as that of the viscosity through our analytical procedure(sections \ref{viscosity_uniform} and \ref{viscosity_gaussian}). Using  \ref{eqn: uni3} and \ref{eqn: norm5}, we get the relations for normalised thermal conductivity for a HS system with Gaussian and uniform distribution respectively.
\begin{align}
\label{norm: l/L_0}
\left(\frac{\kappa}{\kappa_0}\right)_\text{G}&=\frac{2}{(1+\frac{1}{2}\delta^2)(1+\text{erf}(1/\delta))}\\
\label{eqn: l/l_0} 
\left(\frac{\kappa}{\kappa_0}\right)_\text{U}&=\frac{2}{2+\delta^2}
\end{align}
where $\kappa_0$ is the thermal conductivity corresponding to the monodisperse HS fluid. 
\begin{figure}[h]
	\includegraphics[width=1\linewidth]{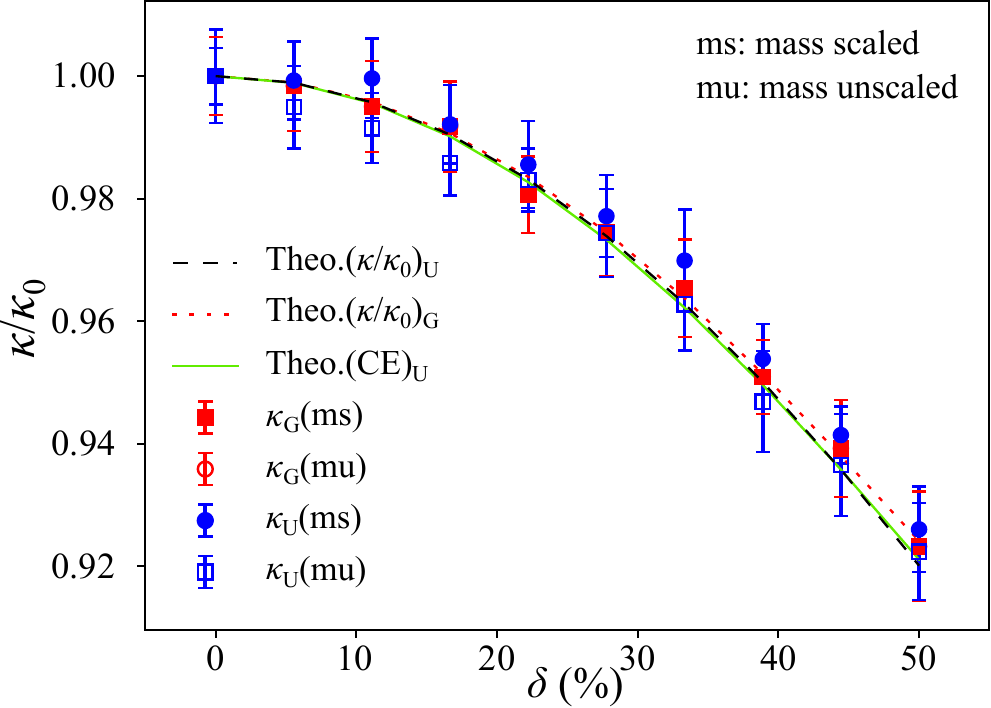}
	\caption{Normalised thermal conductivity of polydisperse HS fluids for different $\delta$ values. Here,Dashed lines represents eqn \ref{eqn: K_U}, dotted line , eqn \ref{eqn: K_G} and the solid line is the Chapman-Enskog method(eqn \ref{eqn: K_CE}). The equation \ref{norm: l/L_0} reduces to equation \ref{eqn: l/l_0} in this range of $\delta$, showing thermal conductivity is independent of the type of particle size distribution(PSD).}
	\label{fig: kappa}
\end{figure}
In the figure~\ref{fig: kappa}, we display the trend of thermal conductivity($\kappa$) for different values of polydispersity of  size polydisperse HS systems with uniform and Gaussian PSDs. The data obtained by MD simulation is fitted with our analytical results (eqns \ref{norm: l/L_0} and \ref{eqn: l/l_0})
The decreasing trend may be understood qualitatively form the behavior of mean free path $\lambda$ based on Kinetic theory which gives direct proportionality relation between $\kappa$ and mean free path $\lambda$. Our analytical solution(eqn \ref{eqn:cs}) predicts increase of collisional cross section(hence, mean free path $\lambda$ decreases)  with polydispersity index($\delta$) for a HS fluid with uniform and Gaussian PSDs. Hence, the decrease in $\kappa$ with $\delta$(the figure \ref{fig: kappa}). Similar to the cases of other transport coefficients, $\kappa$ is also independent of particle size distribution (PSD) type.

\section{Conclusion}
\label{sec: conclusion}
In the present study, we investigate the transport properties of size polydisperse  HS fluids with size polydispersity following Gaussian and uniform particle size distributions(PSDs). For this purpose, we extend the zeroth order Chapman-Enskog solution of Boltzman transport equation to the polydispersity limit as a simpler alternative to the more rigorous method provided by Xu and Stell.\cite{Xu}. In addition, the analytical results are compared  with the data obtained through Molecular Dynamics simulations. we derive simple relations for transport coefficients namely, diffusion constant (eqns \ref{eqn: D/D_0 uni}, \ref{eqn: D/D_0}), shear viscosity (eqns \ref{norm_eta:uni}, \ref{eqn:normfinal}) and thermal conductivity (eqns \ref{eqn: l/l_0}, \ref{norm: l/L_0}) as a function of polydispersity index($\delta$) for HS fluids with size polidispersity following Gaussian and uniform distributions. Also, for the particular case of uniform size distribution, the relation for the shear viscosity is derived using the procedure provided in reference~\cite{Xu} for the case of low polydispersity limit ($\delta<<1$) for a more detailed comparison.\\

Our analytical approach predicts the shear viscosity[eqns \ref{norm_eta:uni}, \ref{eqn:normfinal}] should decrease in quadratic fashion for $\delta<100\%$. The result is found to be accurate in the low polydispersity limit upto $\delta<20\%$ beyond which the shear viscosity obtained by MD simulations increases in agreement with CE solution\cite{Xu} upto $\delta=50\%$. Further, beyond this value of $\delta$, both our analytical result and CE solution show decrease in shear viscosity  although the later decreases at a faster rate. For the case of the Diffusion coefficient($D$) and the thermal conductivity($\kappa$), our analytical soutions as well as CE solution fit well with the MD simulation data  for the entire range of polydispersity index($\delta$) under the study. Both these transport coefficients are found to be monotonic functions of $\delta$. Further, all the transport coefficients under study show dependence on $\delta$ while they are insensitive to the nature of particle size distributions(PSDs).
Our analytical calculation is simple yet accurate for size polydisperse HS fluids with low polydispersity. To sum up, we have provided simple approximate expressions for transport coefficients of polydisperse fluids as a function of polydispersity index $\delta$  which will be useful for the expermental study of real polydisperse fluids. While the simple relations for transport coefficients provided here can be conviniently used in the study of real  fluids. Such model systems can mimick complex heterogeneous environment such as observed in cell cytoplasm where various components of different shape/size are present and colloidal systems in general. 


\begin{acknowledgements}
T.~Premkumar acknowledges fruitful discussions with M.~Premjit, J.~Pame, T.~Vilip, and U.~Somas. T.~Premkumar acknowledges the financial support from NIT Manipur.
\end{acknowledgements}



\renewcommand{\thetable}{A\arabic{table}}   
\renewcommand{\thefigure}{A\arabic{figure}}
\renewcommand{\theequation}{A\arabic{equation}}
\setcounter{figure}{0}
\setcounter{equation}{0}
\appendix

\section{MD Simulation details}
\label{app: simulation-details}

The Lennard-Jones (LJ) pair potential is used  to model the interaction between any two particles $i$ and $j$, represented by\cite{pham}
\begin{equation}
U^{LJ}_{\tiny{_{ ij}}}(r) = 4\epsilon_{ij}\left[\left(\frac{\sigma_{ij}}{r_{ij}}\right)^{12}-\left(\frac{\sigma_{ij}}{r_{ij}}\right)^{6}\right]~,
\label{eqn: LJ-potential}
\end{equation}
where $\sigma_{ij}$ is the arithmetic mean of diameters of the two particles. In order to simulate a hard sphere (HS) system, the cut-off radius for the potential is set at $r_c=1.12\mu$, where the LJ potential is shifted to zero. The energy parameter $\epsilon_{ij}$ is the same for every pair of particles. All the quantities in these MD simulations are expressed in reduced units, length in the units of $\mu$, temperature \textit{T} in $\epsilon/k_B$, pressure in $\epsilon/\mu^3$, and time in $\tau_{LJ}=\sqrt{m\mu^2}/\epsilon$.

The equation of motion of the $i^{th}$ particle is given by the Langevin equation 
\begin{equation}
\label{eqn: Lan}
m_i\frac{\text{d}^2\vec{r_i}}{\text{d}t^2}+\gamma\frac{\text{d}\vec{r_i}}{\text{d}t}=-\frac{\partial U}{\partial\vec{r_i}}+\vec{f_i}(t),
\end{equation}
with $\vec{r_i}$, the position of the $i^{th}$ particle, $\gamma$ is the friction coefficient, \textit{U} is the net pair potential, and $\vec{f_i}(t)$ is a random external force which satisfies the relations: $\langle \vec{f_i}(t)\rangle=0$ and $\left\langle \vec{f_i}^\alpha(t)\vec{f_j}^\beta(t)\right\rangle=2\gamma m_ik_BT\delta_{ij}\delta_{\alpha\beta}\delta(t-t')$, where $\alpha$ and $\beta$ are cartesian components. The friction coefficeint $\gamma=1/\tau_d$, $\tau_d$ being the characteristic viscous damping time which is set to 50 in these simulations. Integration of the equation of motion were performed using velocity-Verlet scheme with time step $\delta t=0.005$.\cite{Allen}. A summary of the systems studied is presented in table~\ref{table: system summary}. 
\begin{table}[h]
	\caption{System summary}
	\centering
	\begin{tabular}{|l|l|}
		\hline
		No. of particles, N & 5000  \\
		\hline
		Size distribution  & (a) Uniform  \\
		(mean value = 1)   & (b) Gaussian \\
		\hline
		Particle mass, $m_i$ & (a) same mass ($m=1$ for all) \\
		& (b) poly-mass (scales with size) \\
		\hline
	\end{tabular}
	\label{table: system summary}
\end{table}

In a simulation box of dimensions $L_x=L_y=L_z=150\mu$ we put $N=5000$ number of particles and then simulations are performed under constant NPT ensemble using Nos\'{e}-Hoover thermostat and barostat\cite{todd} to maintain constant temperature and pressure, respectively. In this study, pressure is set to $P=1$ and temperature, $T=1.5$. 
In MD simulations for polydisperse systems, the result of only one replica of the system is generally not statistically accurate. In order to improve the results, we take a maximum of 30 replicas of each HS system. On comparing the average results of different numbers of replicas for each HS system, it was found that 15 replicas of each system was enough to obtain a set of statistically improved results. Hence, in this study we have tested the reliability of the result by considering an optimum number of replicas of each system  keeping in mind the constraints of precise result, time and computational resources. All the simulations are carried out using open source code LAMMPS.\cite{lammps}


\renewcommand{\thetable}{B\arabic{table}}   
\renewcommand{\thefigure}{B\arabic{figure}}
\renewcommand{\theequation}{B\arabic{equation}}

\section{Transport coefficients}
\label{app: gk-method}

The transport coefficients such as viscosity and thermal conductivity are obtained using Green-Kubo (GK) method\cite{kubo,green,rkubo} which uses the integrals of time autocorrelation functions \cite{vis}. 
According to GK-method, the expression for viscosity coefficient is given as\cite{todd} 
\begin{equation}
\label{eqn: gk}
\eta=\frac{V}{k_BT}\int_0^\infty \text{d}t\langle P_{xy}(0)P_{xy}(t)\rangle,
\end{equation}
where \textit{V} is the volume and $p_{xy}$ is the off-diagonal element of the stress tensor defined as 
\begin{equation}
\label{eqn: p}
P_{xy}=\sum_{i=1}^N\frac{p_i^xp_i^y}{m_i}-\sum_{i>j}^N(x_i-x_j)\frac{\partial u_{ij}}{\partial y_{j}},
\end{equation}
where \textit{i} and \textit{j} particles number indices, $p_i^x$ is the \textit{x} compenent of momentum, $u_{ij}$ is the interaction potential between $i^{th}$ and $j^{th}$ particle, $m_i$ is the mass of the $i^{th}$ particle whose x-component of position vector is $x_i$. 
The thermal conductivity is given by\cite{todd} 
\begin{equation}
\label{eqn: kappa}
\kappa= \frac{V}{3k_BT^2}\int_0^\infty \langle\vec{j}_q(t)\cdot\vec{j}_q(0)\rangle\text{d}t,
\end{equation}  

where $\vec{j}_q$ is the particle flux density. In the following we present a simple analytical approach to study the size-polydisperse system using Boltzmann transport equation. 


\renewcommand{\thetable}{C\arabic{table}}   
\renewcommand{\thefigure}{C\arabic{figure}}
\renewcommand{\theequation}{C\arabic{equation}}

\section{Curve Fitting}
\label{cf}
The MD simulations data are fitted using the analytically obtained expression. In the process, we introduce fitting parameters as listed below:\\

\noindent (a) Diffusion constant:
\begin{eqnarray}
\label{eqn: D_U}
\left(\frac{D}{D_0}\right)_U &=& 1+\frac{A_U}{2}\delta^2\\
\label{eqn: D_G}
\left(\frac{D}{D_0}\right)_G &=& \frac{1}{2}\left(1+\frac{A_G}{2}\delta^2\right)\left(1+\text{erf}(\frac{1}{\delta})\right)\\
\label{eqn: D_C}
\left(\frac{D}{D_0}\right)_{CE} &=& \frac{1}{1+A_{CE}\frac{\delta^2}{2}-\frac{1}{32}\delta^4}
\end{eqnarray}
The fiiting parameters are found are $A_U=1.63,~A_G=1.70$ and $A_{CE}=-1.37$.\\

\noindent (b) Thermal Conductivity:
\begin{eqnarray}
\label{eqn: K_U}
\left(\frac{\kappa}{\kappa_0}\right)_U &=& \frac{2}{2+B_U\delta^2}\\
\label{eqn: K_G}
\left(\frac{\kappa}{\kappa_0}\right)_G &=& \frac{2}{\left(1+B_G\frac{\delta^2}{2}\right)\left(1+\text{erf}\left(\frac{1}{\delta}\right)\right)}\\
\label{eqn: K_CE}
\left(\frac{\kappa}{\kappa_0}\right)_{CE} &=& 1+B_{CE}\frac{\delta^2}{4}-\frac{1}{32}\delta^4
\end{eqnarray}
Here, the fitting parameters are  $B_U=0.69,~B_G=0.68$ and $B_{CE}=-0.63$.\\

\noindent (c) Shear viscosity:
Lastly, we used curve fitting for the casee of CE equation written below for the case of shear viscosity
\begin{equation}
\frac{\eta}{\eta_0}~=~1+C_{CE}\frac{\delta^2}{4}-\frac{1}{32}\delta^4
\end{equation}
and here, $C_{CE}=1$.

	



\begin{thebibliography}{99}
\bibitem{Bird} Bird, R. B., Stewart, W. E., \& Lightfoot, E. N. (1960). Transport phenomena john wiley \& sons. \textit{New York}, 413.

\bibitem{her} Hirschfelder, J. O., Bird, R. B., \& Spotz, E. L. (1948). The transport properties for non‐polar gases. \textit{The Journal of Chemical Physics, 16}(10), 968-981.

\bibitem{Mait}  Gough, D. W., Maitland, G. C., \& Smith, E. B. (1972). The direct determination of intermolecular potential energy functions from gas viscosity measurements. \textit{Molecular Physics}, 24(1), 151-161.

\bibitem{VV} Vesovic, V., \& Wakeham, W. A. (1987). An interpretation of intermolecular pair potentials obtained by inversion for non-spherical systems. \textit{Molecular Physics}, 62(5), 1239-1246.

\bibitem{Duc} Nguyen, D. H., Az\'{e}ma, \'{E}., Radjai, F., \& Sornay, P. (2014). Effect of size polydispersity versus particle shape in dense granular media. \textit{Physical Review E}, 90(1), 012202.

\bibitem{Li} Li, Z. Y., \& Zhang, Z. Q. (2000). Fragility of photonic band gaps in inverse-opal photonic crystals. \textit{Physical Review B}, 62(3), 1516.

\bibitem{Wilding} Wilding, N. B., Sollich, P., \& Buzzacchi, M. (2008). Polydisperse lattice-gas model. \textit{Physical Review E}, 77(1), 011501.

\bibitem{Allar} Allard, M., \& Sargent, E. H. (2004). Impact of polydispersity on light propagation in colloidal photonic crystals. \textit{Applied physics letters}, 85(24), 5887-5889.

\bibitem{Sollich} Sollich, P., \& Wilding, N. B. (2010). Crystalline phases of polydisperse spheres. \textit{Physical review letters}, 104(11), 118302.

\bibitem{Maso} Masoero, E., Del Gado, E., Pellenq, R. M., Ulm, F. J., \& Yip, S. (2012). Nanostructure and nanomechanics of cement: polydisperse colloidal packing. \textit{Physical review letters}, 109(15), 155503.

\bibitem{Xu} Xu, J., \& Stell, G. (1988). Transport properties of polydisperse fluids. I. Shear viscosity of polydisperse hard‐sphere fluids.\textit{ The Journal of chemical physics}, 89(4), 2344-2355.

\bibitem{Dick} Dickinson, E., \& Parker, R. (1985). Polydispersity and the fluid-crystalline phase transition. \textit{Journal de Physique Lettres}, 46(6), 229-232.

\bibitem{Pusey} Pusey, P. N., \& Van Megen, W. (1986). Phase behaviour of concentrated suspensions of nearly hard colloidal spheres. \textit{Nature}, 320(6060), 340-342.

\bibitem{Bol} Bolhuis, P. G., \& Kofke, D. A. (1996). Monte Carlo study of freezing of polydisperse hard spheres. \textit{Physical Review E}, 54(1), 634.

\bibitem{Phan} Phan, S. E., Russel, W. B., Zhu, J., \& Chaikin, P. M. (1998). Effects of polydispersity on hard sphere crystals. \textit{The Journal of chemical physics}, 108(23), 9789-9795.

\bibitem{Barrat} Barrat, J. L., \& Hansen, J. P. (1986). On the stability of polydisperse colloidal crystals. \textit{Journal de physique}, 47(9), 1547-1553.

\bibitem{Rae} McRae, R., \& Haymet, A. D. J. (1988). Freezing of polydisperse hard spheres. \textit{The Journal of chemical physics}, 88(2), 1114-1125.

\bibitem{Bolh} Bolhuis, P. G., \& Kofke, D. A. (1996). Numerical study of freezing in polydisperse colloidal suspensions. \textit{Journal of Physics: Condensed Matter}, 8(47), 9627.

\bibitem{Bart} Bartlett, P., \& Warren, P. B. (1999). Reentrant melting in polydispersed hard spheres. \textit{Physical review letters}, 82(9), 1979.

\bibitem{Wea}Weaire, D., \& Aste, T. (2008). \textit{The pursuit of perfect packing}. CRC Press.

\bibitem{Herr}Herrmann, H. J., Baram, R. M., \& Wackenhut, M. (2003). Searching for the perfect packing. \textit{Physica A: Statistical Mechanics and its Applications}, 330(1-2), 77-82.

\bibitem{Lue} Lue, L. (2005). Collision statistics, thermodynamics, and transport coefficients of hard hyperspheres in three, four, and five dimensions. \textit{The Journal of chemical physics, 122}(4), 04451

\bibitem{Alder} Alder, B. J., Gass, D. M., \& Wainwright, T. E. (1970). Studies in Molecular Dynamics. VIII. The Transport Coefficients for a Hard‐Sphere Fluid. \textit{The Journal of Chemical Physics}, 53(10), 3813-3826.

\bibitem{Sig} Sigurgeirsson, H., \& Heyes, D. M. (2003). Transport coefficients of hard sphere fluids.\textit{ Molecular Physics}, 101(3), 469-482.

\bibitem{Pie} Pieprzyk, S., Bannerman, M. N., Brańka, A. C., Chudak, M., \& Heyes, D. M. (2019). Thermodynamic and dynamical properties of the hard sphere system revisited by molecular dynamics simulation. \textit{Physical Chemistry Chemical Physics}, 21(13), 6886-6899.

\bibitem{Mei} Meier, K., Laesecke, A., \& Kabelac, S. (2001). A molecular dynamics simulation study of the self-diffusion coefficient and viscosity of the Lennard–Jones fluid. \textit{International journal of thermophysics, 22}(1), 161-173.

\bibitem{Mur} Murarka, R. K., \& Bagchi, B. (2003). Diffusion and viscosity in a supercooled polydisperse system. \textit{Physical Review E, 67}(5), 051504.

\bibitem{Fer} Ferziger, J. H., Kaper, H. G., \& Kaper, H. G. (1972). \textit{Mathematical theory of transport processes in gases}. North-Holland.

\bibitem{Thorne} Chapman, S., \& Cowling, T. G. (1970). The Mathematical Theory of Non Uniform Gases, Cambridge Mathematical Library.

\bibitem{Tham} Tham, M. K., \& Gubbins, K. E. (1971). Kinetic theory of multicomponent dense fluid mixtures of rigid spheres. \textit{The Journal of Chemical Physics}, 55(1), 268-279.

\bibitem{van} Van Beijeren, H., \& Ernst, M. H. (1973). The non-linear Enskog-Boltzmann equation. \textit{Physics Letters A}, 43(4), 367-368.

\bibitem{Rief} Reif, F. (2009). \textit{Fundamentals of statistical and thermal physics}. Waveland Press.

\bibitem{sheldon} Ross, S. M. (2020). \textit{Introduction to probability and statistics for engineers and scientists.} Academic press.

\bibitem{chap} Chapman, S., \& Cowling, T. G. (1990). \textit{The mathematical theory of non-uniform gases: an account of the kinetic theory of viscosity, thermal conduction and diffusion in gases}. Cambridge university press.

\bibitem{Julia} Zmpitas, J., \& Gross, J. (2021). Modified Stokes-Einstein Equation for Molecular Self-Diffusion Based on Entropy Scaling. \textit{Industrial \& Engineering Chemistry Research}, 60(11), 4453-4459.

\bibitem{elsana} Elsana, H., Olusanya, T. O., Carr-Wilkinson, J., Darby, S., Faheem, A., \& Elkordy, A. A. (2019). Evaluation of novel cationic gene based liposomes with cyclodextrin prepared by thin film hydration and microfluidic systems. \textit{Scientific reports}, 9(1), 1-17.

\bibitem{JO}Hirschfelder, J. O., Curtiss, C. F., \& Bird, R. B. (1964). Molecular theory of gases and liquids. \textit{Molecular theory of gases and liquids}.

\bibitem{pham} Pham, K.N., Puertas, A.M., Bergenholtz, J., Egelhaaf, S.U., Moussaıd, A., Pusey, P.N., Schofield, A.B., Cates, M.E., Fuchs, M. and Poon, W.C., 2002. Multiple glassy states in a simple model system. {\it Science}, 296(5565), 104-106.

\bibitem{Allen} Allen, M. P. (1987). DJ Tildesley Computer simulation of liquids. Clarendon, Oxford.

\bibitem{todd} Todd, B., \& Daivis, P. (2017). Frontmatter. \textit{In Nonequilibrium Molecular Dynamics: Theory, Algorithms and Applications} (pp. I-Iv). Cambridge: Cambridge University Press.

\bibitem{lammps} Plimpton, S. (1995). Fast parallel algorithms for short-range molecular dynamics. {\it  Journal of computational physics}, {\bf 117}(1), 1-19.

\bibitem{kubo} Kubo, R. (1966). The fluctuation-dissipation theorem. \textit{Reports on progress in physics, 29}(1), 255.

\bibitem{green} Green, M. S. (1954). Markoff random processes and the statistical mechanics of time‐dependent phenomena. II. Irreversible processes in fluids. \textit{The Journal of Chemical Physics, 22}(3), 398-413.

\bibitem{rkubo} Kubo, R. (1957). Statistical-mechanical theory of irreversible processes. I. General theory and simple applications to magnetic and conduction problems. \textit{Journal of the Physical Society of Japan, 12}(6), 570-586.

\bibitem{vis} Viscardy, S., Servantie, J., \& Gaspard, P. (2007). Transport and Helfand moments in the Lennard-Jones fluid. I. Shear viscosity. \textit{The Journal of chemical physics, 126}(18), 184512.


\end{thebibliography}
\end{document}